# NFT-Based Blockchain-Oriented Security Framework for Metaverse Applications

**Khadija Manzoor[1], Umara Noor[1], Zahid Rashid[2]**

[1]Department of Computer Science, Faculty of Computing and Information Technology, International Islamic University, Islamabad, Pakistan
[2]Technology Management Economics and Policy Program, College of Engineering, Seoul National University, 1 Gwanak-Ro, Gwanak-Gu, 08826 Seoul, South Korea
khadijamanzoor63@yahoo.com, umara.zahid@iiu.edu.pk, rashidzahid@snu.ac.kr

**Abstract**

The Metaverse is rapidly evolving, bringing us closer to its imminent reality. However, the widespread adoption of this new automated technology poses significant research challenges in terms of authenticity, integrity, interoperability, and efficiency. These challenges originate from the core technologies underlying the Metaverse and are exacerbated by its complex nature. As a solution to these challenges, this paper presents a novel framework based on Non-Fungible Tokens (NFTs). The framework employs the Proof-of-Stake consensus algorithm, a blockchain-based technology, for data transaction, validation, and resource management. PoS efficiently consume energy and provide a streamlined validation approach instead of resource-intensive mining. This ability makes PoS an ideal candidate for Metaverse applications. By combining NFTs for user authentication and PoS for data integrity, enhanced transaction throughput, and improved scalability, the proposed blockchain mechanism demonstrates noteworthy advantages. Through security analysis, experimental and simulation results, it is established that the NFT-based approach coupled with the PoS algorithm is secure and efficient for Metaverse applications.

1. ## Introduction

The Integration of Artificial Intelligence (AI), the Internet of Things (IoT), and Metaverse technology is transforming physical and virtual worlds into an immersive space. Metaverse technology is the future of the internet where the 2D world is revolutionized into a virtual 3D environment. The individual is represented by an avatar instead of static text or image-based profiles [1]. The technology of Augmented Reality (AR), Virtual Reality (VR), and the Internet is employed to develop meta worlds. The users can enter their own worlds as well as other users' virtual worlds. The users can work, play, and entertain in the Metaverse. The current focus on technological advancements positions the Metaverse as "the successor to the mobile Internet" [1]. In the future, the Metaverse is expected to surpass the capabilities of the traditional Internet, transforming various sectors, including education, healthcare [1], entertainment, e-commerce [2], and smart industries [3]. Blockchain is a shared and ordered database ledger that plays a significant role in storing and organizing data into hashed blocks [4]. It provides a decentralized, auditable, immutable, and transparent environment for data transactions and resource management. There are public, private, and consortium blockchains. There is peer-to-peer communication with varying degrees of decentralization. In distributed blockchain systems, consensus algorithms [5] facilitate the agreement on data values. They determine how data is distributed among blockchain nodes, the decision-making process of each node, and the addition of new verified block transactions to the blockchain. Consensus algorithms are fundamental to different blockchain application schemes as they define the secure validation of individual transactions by shared nodes.

Non-Fungible Tokens (NFTs) are virtual assets that establish ownership of specific physical or virtual items. NFTs can tokenize user-generated content such as artwork, music, videos, and in-game objects (e.g., Roblox [6]). They represent unique and irreplaceable tokens that ensure asset identification and ownership provenance within the blockchain network. By utilizing the blockchain's security and decentralized trust, resource allocation and management in heterogeneous communication networks can be efficiently achieved. The key distinction between NFTs and conventional cryptocurrencies lies in their uniqueness, preventing the value of one NFT from being identical to another. This attribute empowers users to represent their real-world assets digitally on the blockchain and engage in online transactions. The use of smart contracts boosts the development of unlimited NFT-based use cases [7].

The objective of using Proof-of-Stake (PoS) in the Metaverse is that it will provide secure, and transparent banking services, such as stock exchanges and currency exchanges. Unlike traditional mining processes that rely on computational power, PoS replaces complex computations with a validation process. Validators, or nodes, compete based on the banking resources they have staked within the system, rather than computational performance. PoS is an energy-efficient algorithm that selects validators based on the "richest" stakeholder approach. It's important to note that in a distributed blockchain system if a user possesses over 51% of the stake resources, they can potentially control the shared ledger. However, this scenario poses a significant risk, as a malicious user with such control can manipulate or compromise the entire blockchain network [8]. In summary, the existing research highlights the absence of an efficient decentralized mechanism that fully addresses the authenticity, integrity, and interoperability issues associated with data transactions in various metaverse applications. Current blockchain-based schemes also present unresolved challenges in data and resource allocation [5]. To address these concerns, an authenticated NFT-based framework is proposed that ensures data authenticity, integrity, and interoperability. Our approach utilizes NFTs for user authentication and employs the PoS algorithm to provide data integrity, higher transaction throughput, scalability, and a blockchain scheme that positions it as a prime candidate for metaverse applications.

The remaining sections of this research paper are organized as follows: Section II provides a detailed review of related work. Section III presents the proposed NFT-based blockchain-oriented metaverse model. Section IV explains the experimental analysis. Finally, Section V provides the conclusion and future research directions.

## 2. Literature Review

A comprehensive literature review of the key technological trends encompassing the Metaverse, blockchain, the role of blockchain, and NFT in the metaverse is given in the following subsections. The relevance of the consensus algorithm PoS in the context of improved security posture is discussed.

### 2.1 Metaverse Technology

The term Metaverse [1] is a process of integrating the physical world with the virtual world. It allows digital avatars to take out different rich activities like creation, exhibition, entertainment, sociable connections, and trading. It gives the way to figure out an impressive digital world by transforming it into a larger physical world along with the analysis of the Metaverse.

In the paper survey [9], the author's discussed the term Metaverse by explaining how different blockchains including Artificial Intelligence (AI) integrate with it by studying and reviewing the state of art program of studies by the Metaverse contents like digital currencies, applications of AI and blockchain-oriented technologies in the virtual world. Through pervasive wireless networks and the dominant edge computing-based technologies, the Virtual Reality (VR) users within this wireless edge-based metaverse can merge them with the virtual environment by accessing those VR services as offered by multiple VR providers.

The authors in [10] proposed an Incentive mechanism scheme for learning-based VR services within the Metaverse. They used a non-panoramic virtual reality-based environment to shorten the data movement and computation utilization. They applied Structural Similarity (SSIM) and Video Multi-Method Assessment Fusion (VMAF) to jointly define the perception background of the VR users. After this, they have also proposed the method named double Dutch auction which promotes the trading, matching, and pricing processes through multiple VR service providers and the users asynchronously. Additionally, to further increase this proposed learning mechanism efficiency; they have trained the auctioneer without prior auction knowledge in the act of the learning agent through Deep Reinforcement Learning (DRL) scheme.

### 2.2 Blockchain Technology

Initially, blockchain was primarily recognized as a method for storing commercial data and serving as a distributed channel database for multiple commercial transactions. However, this trend has now expanded to various domains. For instance, blockchains are being applied in the Internet of Things (IoT) environment across different fields such as healthcare services, leveraging technologies like the Internet of Skills [11]. Other areas where blockchain is being employed include automated manufacturing processes [12], secure data aggregation [13], mixed reality content sharding [13], and COVID pandemic monitoring [15].

Furthermore, in [16], the authors proposed an approach that combines blockchain and smart contracts for service management in IoT. They demonstrated the potential of decentralized consensus mechanisms based on blockchain to ensure secure interoperability among various sensors and actuators operating within IoT systems [17].

In the research paper [18], the authors proposed the use of the Inter Planetary File System (IPFS) mechanism to mitigate decentralized database risks. IPFS is commonly employed to achieve multiple system backups and ensure secure data content addressing. Unlike centralized storage solutions that rely on a central server, the IPFS mechanism operates on a network of nodes, eliminating concerns regarding data invalidation and tampering.

In [19], the authors highlighted that while off-chain storage systems offer increased storage capacity, the scalability of these systems still needs improvement to meet the growing demands of NFT-based storage structures in the Metaverse. To address this, they proposed a multiple blockchain-based storage system that enhances the storage capacity of the blockchain for NFTs. This system introduces multiple blockchains that can be combined to create a unified NFT-based storage platform, utilizing both blockchain and off-chain storage resources. However, due to the limitations of the underlying blockchain platforms mentioned, existing NFT schemes lack interoperability, resulting in separate implementations.

### 2.3 Role of Blockchain in Metaverse

The synchronization between the physical and virtual worlds heavily relies on the use of blockchain-based data processing algorithms to provide corresponding views and create control instructions in both directions. However, future blockchain-based metaverse applications will face various security and robustness challenges against external attacks [20].

In their research paper [21], the authors proposed MetaChain, a unique blockchain-based framework designed to efficiently manage Metaverse applications. The MetaChain framework leverages the advantages of blockchain-based sharding technologies to establish smart and trustworthy interactions between Metaverse Service Providers (MSPs) and Metaverse Users (MUs). They also presented a novel blockchain-based sharding scheme to enhance MetaChain's performance and utilize MUs' resources for Metaverse applications. The authors conducted a Stackelberg game theoretical analysis, examining MUs' behaviors and developing an economic model that allows MSPs to incentivize and allocate MUs' resources based on the demands of Metaverse applications.

In [22], the authors proposed a Public Key Infrastructure (PKI) system that generates a securely closed domain and recognizes certificates only within that domain, such as in scientific systems, financial systems, and 5G networks. They introduced an efficient privacy-preserving cross-domain authentication scheme called XAuth, which was integrated with the existing PKI system and a Certificate Transparency (CT) system. Additionally, they designed a lightweight correctness verification protocol using multiple Merkle Hash Trees to provide fast responses. To protect users' privacy, they proposed an anonymous authentication protocol for cross-domain authentication. The authors conducted a security analysis and presented experimental results to demonstrate the performance efficiency of the proposed XAuth scheme.

Furthermore, [23] discusses how wearable devices can generate low-throughput sensory data to capture user actions and gestures in virtual reality (VR), while VR devices can deliver corresponding high-throughput 3D video streams based on user movements. This capability offers new user experiences in various industries such as remote surgeries, machinery maintenance, holographic telepresence, and autonomous driving [24].

### 2.4 NFT in Metaverse and Blockchain

NFTs have become a prominent component in the development of metaverse solutions within the blockchain industry. Currently, NFTs are widely utilized in the gaming and entertainment sectors, where they serve as unique digital assets in various gaming and metaverse applications [25].

In [18], the authors integrated NFTs into the metaverse to create new digital content for business purposes. Additionally, [26] highlights the longer-term impact of NFTs in the context of blockchain, AI, and industry, emphasizing their potential to securely and trustfully digitize real, physical, or virtual assets.

NFTs play a crucial role in confirming the uniqueness of digital assets by securely recording transaction records on the blockchain. Each NFT possesses a unique identification code that verifies ownership of the associated digital assets and serves as a reference for future transactions [19]. NFTs are especially utilized to commemorate special moments and acquire virtual property.

In [6], for each blockchain, separate NFTs are implemented for data flow application. These NFTs were governed by smart contracts established between user devices and mobile network operators (MNOs) operating on the proposed blockchain infrastructure.

### 2.5 Proof-of-Stake (POS) Algorithm

Frauenthaler et al. [27] introduced a blockchain-based technique using PoS to reduce the cost of block validation and prioritize the validation of block headers when necessary. The proposed mechanism aims to enhance interoperability between Ethereum-based blockchains. It involves establishing transactions based on the number and age of coins used by miner nodes, leading to optimal results for a set of proposed transactions.

In [28], the authors proposed a stochastic resource allocation scheme called SORAS, which is based on stochastic integer programming. This scheme optimizes resource allocations for Metaverse users (MUs) while minimizing costs for digital service providers within the Metaverse.

Additionally, in [29], the authors proposed a distributed cryptocurrency system that utilizes a decentralized cryptocurrency trading protocol. This protocol employs two consensus mechanisms, Proof of Work (PoW) and Proof of Deposit (PoD), to select trustworthy users for a validation committee. The number of participants impacts the platform provision and the execution cost instead of the number of transactions by players.

In view of the above, there is a research gap in the context of data transactions for various blockchain-based Metaverse applications. To address the problems of authentication, integrity, and resource allocation in the Metaverse, efficient and decentralized techniques are required.

## 3. NFT-based Blockchain-Oriented Metaverse

A blockchain-oriented framework that employs NFTs and the PoS algorithm is proposed as a solution to the problems of authentication, integrity, and resource allocation in Metaverse applications. The decentralization solves the issues of data authentication and, integrity.

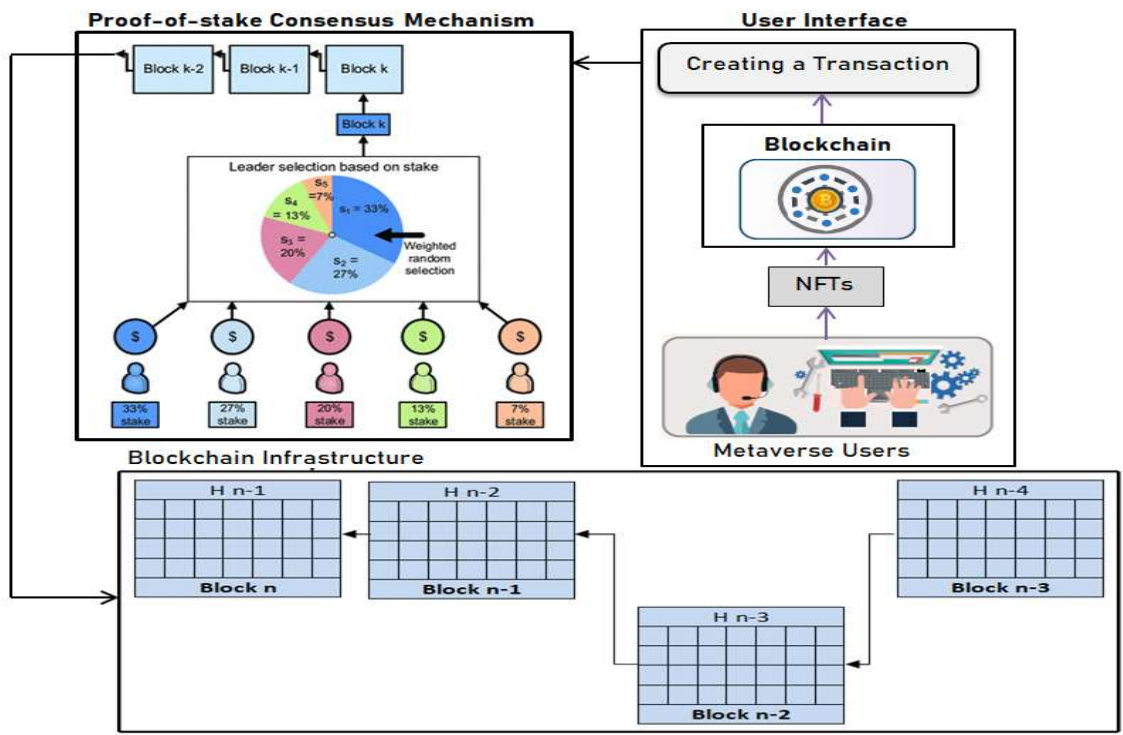

*Figure 1. NFT-based Blockchain-oriented Metaverse*

For every user a unique NFT that serves as digital identity is generated. The purpose is user authentication. The user can do transactions using the unique NFT in the Metaversee application. Now for the purpose of integrity, and interoperability of user's transactions, PoS consensus algorithm is employed. PoS's is an energy-efficient consensus algorithm that replaces the traditional complex mining process of blockchain with an efficient validation process. These improvements provide an eminent high transaction throughput and scalability in the Metaverse environment.

We propose a decentralized model where each user in the blockchain-based application system has its own block transaction and unique identifier. These unique identifiers can be recognized by all registered client users within the Metaverse. The implemented system model, showcasing the interaction between the blockchain, Metaverse users, their identifications, and transaction management, is illustrated in Figure 1.

To initiate a transaction and prepare it for the validation process, Metaverse client users must first log into the blockchain-based system using their unique NFTs, which serve as their usernames and passwords. Following this, the client users are required to share their resources to validate the created transaction blocks. The consensus algorithm PoS is applied, designating a leader known as the validator. The validator role is granted only to those clients who have shared more resources during the transaction process. Subsequently, the validator node in the Metaverse system is given access control over the block transactions.

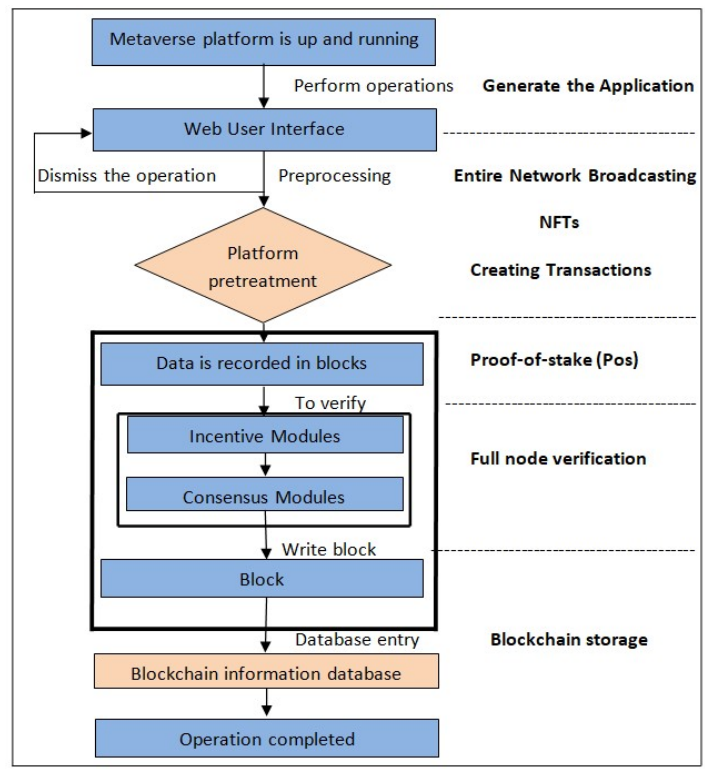

*Figure 2. Operational flowchart of the NFT-based blockchain-oriented framework*

The proposed operational flowchart for the metaverse system, depicted in Figure 2, outlines the process of creating transactions, network data communication, incentive mechanisms, consensus modules, and blockchain storage. To initiate a transaction, client users must first generate their own applications within the blockchain-based Metaverse environment. Once their applications are created, users are assigned unique NFTs, which enable them to create block transactions.

To validate these block transactions, client users are required to share sufficient resources within the blockchain-based Metaverse system. The shared resource records are maintained, and the PoS consensus algorithm is applied. The user who contributes the most resources is granted the authority by the PoS consensus algorithm to become a validator for transactions within the blockchain-based Metaverse system.

All block transactions, including those validated by the selected validator and subsequent blocks, are stored in a database known as the blockchain information database. Each transaction block within this database includes its own data as well as the hash value of the previous block.

## 4. Experimental Analysis

The experimental analysis is performed using the Ethereum blockchain. It is augmented with NFTs and the PoS consensus algorithm. To verify the users data authenticity, integrity and interoperability, a practical scenario is potentially checked in which the Ethereum blockchain is implemented with NFTs and consensus algorithm Proof-of-Stake (PoS) in Metaverse. This framework considers the implementation of the hybrid transaction management system in which POS user's with their unique NFTs vote to validate the blocks that shares their more resources during the transactions process carried within the metaverse system. In order to become a validator, participant's users have to share their maximum resources within the carried blocks transactions. After choosing the validator, the process of serialization is completed by setting the length of every block transaction to 50 bytes. Then the system test and verify the blockchain-enabled PoS consensus algorithm by checking the throughput values (TPs) and scalability.

### 4.1 Throughput Analysis

In order to measure the performance metric of distributed systems, the transaction throughput value is calculated. System transaction throughput values are calculated by the total number of concurrent block transactions and the number of transactions made per second. The amount of time used by the clients and server to validate a transaction is the time spent on a transaction. Overall system performance depends upon the transactions processing abilities to become a validator. The throughput values of our distributed Metaverse system are calculated based on the number of transaction blocks processed within a given time period, using equation 1:

$$\text{Throughput/sec} = \frac{\text{Sum of Transactions}}{\Delta t} \qquad (1)$$

Where $\Delta t$ is the total time taken for transaction processing, the Sum of transactions is the total number of completed transactions within $\Delta t$, and Throughput per second is the TPS value for the given time period. To accurately gather test data, we calculated the total number of transactions and their corresponding TPS values within a one-minute time period.

A sequential process is employed in the construction of the client transactions. The value of each service transaction is fixed to 50 bytes. To ensure the success of concurrent transactions the tests are conducted.

After completing the test process, we uniformly calculated the TPS values for each transaction within the specified time period in the blockchain-based Metaverse system. The results of the tests are displayed in Figure 3. As shown in Figure 3, our proposed algorithm, which combines NFT-based blockchain with the PoS consensus algorithm, outperforms a simple blockchain system in the Metaverse.

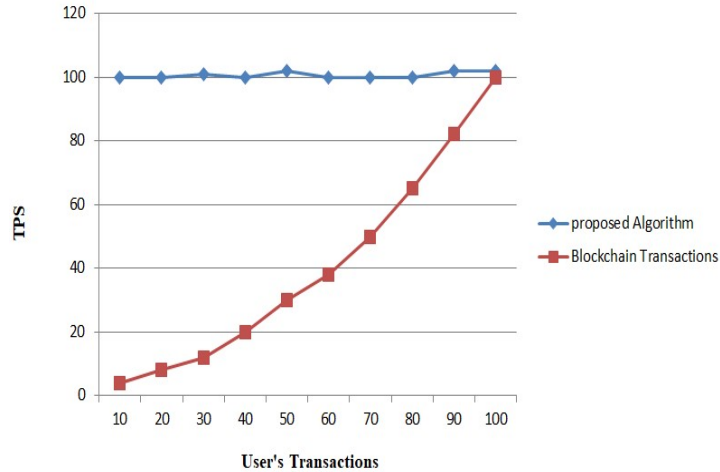

*Figure 3. Improved Algorithm TPS value with user's transactions*

The value of TPS remains stable and almost constant even for an increased number of node transactions in the PoS-based metaverse system. However, in non PoS-based blockchain metaverse, the TPS values fluctuated when the number of node transactions are either increased or increased.

### 4.2 Scalability Analysis

In our distributed blockchain-based Metaverse system, scalability is achieved through vertical and horizontal expansions. Vertical expansion focuses on enhancing system performance by increasing the number of transactions made by client users within the blockchain-based Metaverse system. To evaluate scalability, we conducted tests by dynamically adjusting the number of block transactions and analyzing the system performance using the PoS consensus algorithm. Specifically, we measured the Transactions Per Second (TPS) values within a one-minute timeframe.

During the scalability test, we observed that the TPS values remained relatively constant regardless of whether we increased or decreased the number of transactions within the specified time period.

## 5. Conclusion

As the Metaverse technologies are continuing to evolve, the security requirements of data transactions are getting interest by the research community. One such solution is to employ consensus algorithms of blockchain technology. In this research paper, a novel blockchain-based solution is presented that integrates NFTs and the PoS consensus algorithm for authentication, integrity, and scalability for metaverse-based applications. This combination emerges as the first of its kind that results in high throughput and an excellent scalability for blockcain based metaverse applications. This makes it an ideal choice for a wide range of Metaverse applications. For future research, it is possible to extend the proposed framework algorithm to achieve even higher levels of security and scalability within the Metaverse system. Additionally, the framework can be adapted to support other sectors beyond the metaverse, enabling its application in diverse domains.